\documentclass[%
 reprint,
superscriptaddress,
 amsmath,amssymb,
 aps,
]{revtex4-1}

\usepackage{amsthm}
\usepackage{graphicx}
\usepackage{dcolumn}
\usepackage{bm}
\usepackage{mathrsfs}
\usepackage{dsfont}

\begin{document}

\preprint{APS/123-ART}

\title{Classical versus quantum completeness}
 \author{Stefan Hofmann}%
 \email{stefan.hofmann@physik.uni-muenchen.de}
\affiliation{Arnold Sommerfeld Center for Theoretical Physics, Theresienstra{\ss}e 37, 80333 M\"unchen\\}
\author{Marc Schneider}%
 \email{marc.schneider@physik.uni-muenchen.de}
 \affiliation{Arnold Sommerfeld Center for Theoretical Physics, Theresienstra{\ss}e 37, 80333 M\"unchen\\}



\date{\today}

\begin{abstract}
The notion of quantum-mechanical completeness is adapted to situations 
where the only adequate description is in terms of quantum field theory
in curved space-times. 
It is then shown that Schwarzschild black holes, although 
geodesically incomplete, are quantum complete.

\begin{description}
\item[PACS numbers]
03.65.-w, 03.65.Db, 03.70.+k, 04.20.Dw, 11.10.-z, 11.10.Ef.
\end{description}
\end{abstract}

\pacs{Valid PACS numbers appear here}
\keywords{Suggested keywords}
\maketitle


\section{\label{sec:level1}Introduction}
Completeness is a very important concept in classical and quantum physics. 
The classical motion on a half-line
is called complete at the end point 
if there are no initial conditions such that the trajectory runs off to the end point
in a finite time. If the potential satisfies certain regularity conditions, 
then the classical motion is complete at the end point
if and only if the potential grows unbounded from above near the end point \cite{ree74}.
In general relativity, a space-time is called geodesically complete, if every maximal geodesic
is defined on the entire real line. If the space-time is timelike or null geodesic incomplete,
it is said to be singular \cite{hawk70}. The physical relevance of this geometrical notion is provided upon
identifying geodesics with trajectories of free test particles.
In quantum mechanics on a half-line, 
a time-independent potential is called quantum-mechanically complete \cite{ree74},
if the associated Hamiltonian is essentially self-adjoint on the space of $C^\infty$-functions
of compact support on the half-line with the origin excluded. 

Horowitz and Marolf \cite{hor95} showed that there are geodesically incomplete static space-times,
with timelike curvature singularities, which are quantum-mechanically complete.
Their work stimulated a lot of research concerning
geodesically incomplete but quantum-mechanically complete spacetimes, 
e.g.~\cite{ish99,ish03,kon04}.
As a working analogue, they suggested the nonrelativistic hydrogen atom.
The classical motion of the electron in the Coulomb potential 
is incomplete at the origin, because the potential is bounded from above 
near the origin and thus the origin can be reached by the electron in a finite time. 
The Coulomb potential is, however, quantum-mechanically complete
when probed by the nonrelativistic bound-state electron. 
In other words, the classical singularity of the Coulomb potential 
is not reflected in any observable related to the bound-state electron.

Quantum field theory in a static, globally hyperbolic space-time
allows to define a consistent quantum theory for a single relativistic particle,
where the energy of each one-particle state is equal to that of the corresponding
classical field \cite{ash75}. 
Horowitz and Marolf  \cite{hor95} showed that this is still the case for certain
static space-times with timelike singularities. 
Their result is based on a work by Wald \cite{wald78, wald80}, 
who proved that the problem of defining 
the evolution of a Klein-Gordon scalar field in an arbitrary static space-time
(with arbitrary singularities consistent with statics) can be reformulated as
the problem of constructing self-adjoint extensions of the spatial part 
of the wave operator.

For a general time-dependent space-time, there is no consistent quantum theory
of a single free particle, and the only adequate description is in terms 
of quantum field theory. This requires to study the 
evolution of classical test fields in a singular space-time. 
In static space-times, the evolution of quantum fields is unitary and represents
an endomorphism of the physical Hilbert space. In particular, unitarity preserves
state normalisation. 
If dynamical space-times are treated as
external backgrounds, the quantum theory does not require a unitary evolution \cite{helf96}.
Therefore, the notion of 
quantum-mechanical completeness needs to be adapted to include this case.

In discussing geodesic completeness, it usually suffices to consider geodesics
defined on $(0,t_0]$, right end points can be treated similarly. A convenient
topological criterion for the inextendibility of a geodesic $\gamma(t)\;, t\in (0,t_0]$
is the following: There is a parameter sequence $\{t_n\}\rightarrow 0$
such that $\{\gamma(t_n)\}$ does not converge. 
As is well known, geodesic 
parametrisations have geometric significance. If a curve has a 
reparametrisation as a geodesic, it is called a pregeodesic.
In particular, any spacelike or timelike curve is pregeodesic if and only if
its reparametrisation by its arc-length yields a geodesic.
A spacelike or timelike pregeodesic $\alpha(t)\;, t\in(0,t_0]$ is complete
(to the left) if and only if it has infinite length \cite{kob63}. 

We call a globally hyperbolic space-time quantum complete (to the left)
with respect to a free field theory, if the Schr\"odinger wave
functional of the free test fields can be normalised at the initial time $t_0$,
and if the normalisation is bounded from above by its initial value
for any $t\in(0,t_0)$. Note that neglecting backreaction is no severe 
restriction here, since if backreaction becomes important, the question
whether a previously given space-time is quantum complete 
becomes obsolete. 

Intuitively, the notion of quantum complete space-times refers to the following:
A space-time background can be considered as an external source coupled to 
quantum fields. This coupling is consistent provided that the norm of the 
vacuum-to-vacuum transition amplitude does not exceed the corresponding norm in 
Minkowski space-time, i.e.~unity. 
There is no conceptual problem if the transition is less probable than in Minkowski
space-time, albeit the evolution is then nonunitary. In this case, the ground state is not 
persistent, but its norm is reduced by transferring probability to the space-time
background, which is not resolved into dynamical degrees of freedom. 
If, in contrast, the transition is more probable, then unitarity is violated in such a way that 
the quantum theory becomes meaningless. 
This intuition is stated more precisely in our definition of quantum completeness, which is
based on the functional Schr\"odinger approach to quantum field theory. As 
opposed to an asymptotic framework pertinent to a scattering description,
the functional Schr\"odinger approach allows to analyse unitarity violations occurring 
during a finite amount of time, and, in particular, during the time interval $(0,t_0]$. 

For a Schwarzschild black hole, a Cauchy hypersurface is given by
$\{t_0\}\times\mathbb{R}\times S^2$, where $t_0\in(0,2M)$ and $M$
denotes the black hole mass. It follows that the black hole interior
is globally hyperbolic and foliated by smooth spacelike Cauchy 
hypersurfaces \cite{wald91}. 
The purpose of this article is to show that the interior
of a Schwarzschild black hole is quantum complete, albeit it is geodesically incomplete.

\section{Set-up}
We briefly review the functional Schr\"odinger formulation for
quantum field theory in generic space-times, when backreaction
can be neglected (for a detailed discussion in Minkowski space-time, see \cite{hat91}). 
This formulation will prove
to be efficient for investigating qualitative features such as
the stability of ground states and the quantum (in)completeness
of generic space-times.

Due to a theorem by Geroch \cite{ger70}, a globally hyperbolic space-time 
is diffeomorphic to $\mathbb{R}\times\Sigma$,
and foliates into hypersurfaces $\Sigma_t\;, t\in\mathbb{R}$.
In the $(1+3)$-split formulation, the classical theory for a free scalar field of mass 
$m$ is
given by the Hamiltonian 
\begin{eqnarray}
	H
	=
	\int_{\Sigma_t}\mbox{d}\mu(x) \left(N_{_\perp} \mathcal{H}^{\perp}
	+ N_{{\shortparallel}}^{\; a}\mathcal{H}_{\; a}^{\shortparallel}\right)
	\;. 
\end{eqnarray}	
Here, $\mbox{d}\mu(x)\equiv {\rm d}^3x \sqrt{{\rm det}(q)}$, 
with $q$ denoting the spatial part of the metric,
$\mathcal{H}_{\; a}^{\shortparallel}=\pi\partial_a\Phi/\sqrt{{\rm det}(q)}$, 
where $\pi=\partial\mathcal{L}/\partial\dot{\Phi}
=\sqrt{\mbox{det}(q)}(\partial_t\Phi-N_{\shortparallel}^a\partial_a\Phi)/N_\perp$
denotes the canonical momentum field, and
\begin{eqnarray}
	\mathcal{H}^{\perp}&=&\tfrac{1}{2}
	\left[\tfrac{1}{{\rm{det}}(q)}\pi^2+q^{ab}\partial_a\Phi\partial_b\Phi
 	+ \left(m^2+\zeta R\right) \Phi^2\right] \; .
\end{eqnarray}	
Here, all tensors are pulled-back to the hypersurface $\Sigma_t$,
and $\zeta$ is a numerical factor representing the 
non-minimal coupling to gravity.
Adapting the space-time coordinates to the foliation, $N_\shortparallel=0$ and
$N_\perp=\sqrt{-g_{tt}}$. 

Each hypersurface $\Sigma_t$ is equipped with a Fock space. 
In the Schr\"odinger representation, the basis of this Fock space
is constructed from the time-independent operator $\Phi(x)$.
Its spectrum contains the classical fields $\phi(x)$ as eigenvalues \cite{hat91}. 
The $\phi$-representation of an arbitrary state $|\Psi\rangle$ in the Fock space
is a (nonlinear) wave functional $\Psi[\phi](t)$. For the momentum field $\pi$ 
canonically conjugated
to $\Phi$, the functional version of the quantisation prescription is 
given by $\pi(x)\rightarrow -{\rm i}\delta/\delta\Phi(x)$. 

$\Psi[\phi](t)$ satisfies a functional generalisation of the Schr\"odinger equation,
\begin{eqnarray}
	\label{fSch}
	{\rm i}\partial_t \Psi[\phi](t) &=& H[\Phi](t) \; \Psi[\phi](t)
	\; , \\
	H[\Phi](t) &=&
	\int_{\Sigma_t} \mbox{d}\mu(x) \; \mathcal{H}(\Phi(x);t,x) \; ,
\end{eqnarray}	
where
$H[\Phi](t)$ denotes an operator valued functional constructed from 
the Hamilton density
\begin{eqnarray}
	\mathcal{H}
	&=&
	\tfrac{1}{2} \Big[
	\tfrac{\sqrt{-g_{tt}}}{{\rm det}(q)}
	\tfrac{\delta^2}{\delta\Phi^2}
	+ q^{ab}\partial_a\Phi\partial_b\Phi 
	+ \left(m^2+\zeta R\right) \Phi^2
	\Big] \; .
\end{eqnarray}	
Note that any explicit time dependence 
is due to the space-time geometry, which can be thought of
as an external source nonminimally coupled to the 
quantum field.

Wave functionals are normalised in the usual sense,
\begin{eqnarray}
	\|\Psi\|^2(t)
	= \int {\rm D}\phi \; \Psi^*[\phi](t) \Psi[\phi](t)
	\; ,
\end{eqnarray}	
where ${\rm D}\phi$ denotes the measure over all field configurations 
in $\Sigma_t$. Stability of the state populated with $\phi(x)$
requires that the norm of the wave functional is time-independent. 
This corresponds to a unitary evolution. 

On a dynamical space-time,
considered as an external background, however, 
the evolution is not required
to be unitary, i.e.~$H[\Phi](t)$ needs not be a self-adjoint
operator on the space of wave functionals. Intuitively, probability 
can be lost to the background (like for dissipative systems when
the interaction causing the friction is not fully resolved in the 
participating degrees of freedom). Consistency of the dynamics
is more subtle in this case. Let $\|\Psi[\phi]\|^2(t_0)$ denote 
the probability density (with respect to the space of field configurations)
at the initial hypersurface, and consider the interval $(0,t_0]$ 
with zero marking the left end point. We call the evolution consistent,
even if it violates unitarity, provided that 
$ \|\Psi[\phi]\|^2(t)\le \|\Psi[\phi]\|^2(t_0)\;, \forall t\in(0,t_0)$.
Intuitively, probability must not be gained from a background which
is not resolved in dynamical degrees of freedom. If the above consistency
relation is violated, then backreaction effects are relevant, and the 
original space-time geometry is obsolete.


For the time-dependent ground state, a generalised Gaussian ansatz
is motivated following the example of the harmonic oscillator in quantum mechanics:
\begin{eqnarray}
	\label{ansatz}
	\Psi^{(0)}[\phi](t)
	&=& N^{(0)}(t) \; \mathcal{G}^{(0)}[\phi](t) \; , \\
	\mathcal{G}^{(0)}[\phi](t)
	&=&
	\exp{\left[-\tfrac{1}{2}\int_{\Sigma_t}\mbox{d}\mu(x) \mbox{d}\mu(y)
	\phi(x) K(x,y,t) \phi(y) \right]}
	\; . \nonumber
\end{eqnarray}
Substituting the ansatz (\ref{ansatz}) in the 
functional Schr\"odinger equation (\ref{fSch}) gives 
for the $\phi$-independent factor $N^{(0)}(t)$ an evolution equation that
can be directly integrated,
\begin{eqnarray}
	\label{N0}
	N^{(0)}(t) &=&
	N_0\exp\!{\left[-\tfrac{\rm i}{2}
	\int_{t_0}^t\!\!{\rm d}t^\prime\!\! \int_{{\Sigma_{t^\prime}}}
	\!\!\!\sqrt{-g_{tt}}\mbox{d}\mu(z) K(z,z,t^\prime)
	\right]}
	\; \;\;\; 
\end{eqnarray}	
while the evolution for the kernel $K(x,y,t)$ is described by 
a $\phi$-dependent nonlinear integro-differential equation,
\begin{eqnarray}
	\label{kernel}
	&& \frac{{\rm i}\partial_t \left[\sqrt{{\rm det}(q)}(x)\sqrt{{\rm det}(q)}(y)
	K(x,y,t)\right]}{\sqrt{{\rm det}(q)}(x)\sqrt{{\rm det}(q)}(y)}
	= 
	\nonumber \\
	&& 
	\int_{\Sigma_t} \sqrt{-g_{tt}}(z)\mbox{d}\mu(z)\;
	K(x,z,t)K(z,y,t)+
	\nonumber \\
	&&
	+\sqrt{-g_{tt}}(x) \left(\Delta - m^2 -\zeta R\right) \delta^{(3)}(x,y)
	\; .
\end{eqnarray}	 
The spatial part of the Laplace-Beltrami 
operator is defined as $\Delta\equiv\partial_a[\sqrt{\mbox{det}(q)}q^{ab}\partial_b]/\sqrt{\mbox{det}(q)}$,
and we use the following convention for the Dirac distribution:
$\sqrt{{\rm det}(q)}(x)\delta^{(3)}(x,y)\equiv\delta^{(3)}(x-y)$.

\section{Calculation}
In this section, we specialise to the interior of Schwarzschild black holes.
In the usual Schwarzschild coordinate neighbourhood, the Schwarzschild 
function is given by $h(\tau)=(2-\tau)/\tau$, where $\tau\equiv 2t/r_{\rm g}$
is dimensionless, and $r_{\rm g}\equiv 2M$ denotes the Schwarzschild
radius ($G_{\rm N}\equiv 1$). The warped product line element
for the Schwarzschild black hole becomes
\begin{eqnarray}
	g&=& - h^{-1}(\tau) {\rm d}t^2 + h(\tau) {\rm d}r^2 + 
	(\tau r_{\rm g})^2{\rm d}\mathfrak{s}^2/4
	\; ,
\end{eqnarray}
where by this normalisation, in each rest-space $t=$constant, the surface
$r=$constant has the induced line element 
$(\tau r_{\rm g})^2 {\rm d}\mathfrak{s}^2/4$, 
and is 
thus the two-sphere of radius $\tau r_{\rm g}/2$ 
with Gaussian curvature $4/(\tau r_{\rm g})^2$ and area $\pi (\tau r_{\rm g})^2$.
In this parametrisation, the geometry is incomplete to the left, since tidal forces 
approach infinity along inextendible timelike geodesics 
as $\tau\rightarrow 0$.

Since the Schwarzschild space-time is spherically symmetric, the kernel $K$ introduced
in (\ref{ansatz}) is a function $K(x-y,\tau)$. Our convention for Fourier transforms is 
\begin{eqnarray}
	K(z,\tau)
	&=&
	\int \frac{{\rm d}^3k}{(2\pi)^3} \;
	\exp{\left({\rm i q(k,z)}\right)} \; \widehat{K}(k,\tau)
	\; ,
\end{eqnarray}	
with $q(k,z)\equiv q^a_{\; b} k_a z^b$. The Fourier amplitudes 
$\widehat{K}\equiv \widetilde{K}/{\rm det}(q)$ satisfy 
a Riccati equation,
\begin{eqnarray}
	\label{RicattiKernel}
	&&{\rm i}\partial_\tau  \widetilde{K}(k,\tau)
	= \nonumber \\
	&&\sqrt{{\rm det}(g)} \tfrac{r_{\rm g}}{2}\left[
	\left({\rm det}(q)\right)^{-1}  \widetilde{K}^2(k,\tau)
	-  \Omega^2(k,\tau)
	\right]
	\; .
\end{eqnarray}	
The inhomogeneous contribution $\Omega^2(k,\tau)\equiv q^{ab}k_a k_b + m^2$
is just the dispersion relation of the free fields. 

The kernel can alternatively be described 
as follows. Suppose $\varphi(x^\prime,t^\prime)$ is a solution of the equation of motion 
for the free fields. It is related to a solution at a later time $t>t^\prime$ by
Huygens's principle \cite{bak87,ree79},
\begin{eqnarray}
	&&\varphi(x,t)
	=
	\nonumber \\
	&& \int_{t_0}^t {\rm d}t^\prime h^{-1/2}
	\int_{\Sigma_{t^\prime}} \!\!\sqrt{{\rm det}(q)}
	{\rm i} K(x-x^\prime, t^\prime) \varphi(x^\prime,t^\prime)\mbox{d}^3x^{\prime}
	\; .\label{kernhuy}
\end{eqnarray}	
Indeed, a kernel fulfilling Huygens' principle for the time-dependent
fields $\varphi$ is a solution of the kernel equation (\ref{kernel}).
Moreover, 
\begin{eqnarray}
	\label{eom}
	\left(\Box -\Omega^2(k,\tau)\right) 
	\widehat{\varphi}(k,\tau)
	= 0
	\;.
\end{eqnarray}	
Of course, from the solutions of (\ref{eom}) the kernel can 
be calculated directly,
\begin{eqnarray}
	\label{altkernel}
	\widehat{K}(k,\tau)
	&=&
	\tfrac{-{\rm i}}{\sqrt{|{\rm det}(g)|}}\; 
	\partial_t \; {\rm ln}\, \widehat{\varphi}(k,\tau)
	\; ,
\end{eqnarray}	
but it should be clear that this is a less efficient approach
than solving the kernel equation (\ref{RicattiKernel}). With the 
kernel representation (\ref{altkernel}), however, it is straightforward 
to show that the time dependence of $\|\Psi^{(0)}\|$ is 
not fictitious, even without solving (\ref{eom}).
Using (\ref{altkernel}) in (\ref{N0}), we find
\begin{eqnarray}
	\label{normN}
	\left|N^{(0)}(\tau)\right|^2\!\!
	= \left|N_0\right|^2 \!
	\exp\!{\left(-\tfrac{{\rm v(\Sigma)}}{2}
	\int\tfrac{{\rm d}^3k}{(2\pi)^3} \; 
	{\rm ln}\left|\tfrac{\widehat{\varphi}(k,\tau)}{\widehat{\varphi}(k,\tau_0)}\right|^2
	\right)} ,\;\;\;\;
	\end{eqnarray}	
where $v(\Sigma)$ denotes the time-independent coordinate volume of the
hypersurfaces. Furthermore, 
\begin{eqnarray}
	\label{normG}
\left\|\mathcal{G}^{(0)}\right\|^2(\tau)
	= \left({\rm Det}
	\left(\tfrac{{\rm det}(q)}{\sqrt{{\rm det}(g)}} \tfrac{\rm i}{2}
	\tfrac{W(\widehat{\varphi},\widehat{\varphi}^*)}
	{|\widehat{\varphi}|^2}
	\right) \right)^{-1/2}\; ,
\end{eqnarray}		
with 
$W(\widehat{\varphi},\widehat{\varphi}^*)\equiv
\widehat{\varphi}\overleftrightarrow{\partial_t}\widehat{\varphi}^*$ denoting
the Wronskian of the solution and its complex conjugate, and Det is
the functional determinant. From this result, we can
draw two important immediate conclusions. First, for Friedman space-times, 
Abel's differential equation identity \cite{bohn01}
gives that
$\sqrt{|{\rm det}(g)|}W(\widehat{\varphi},\widehat{\varphi}^*)$
is time-independent. As a consequence,  $\|\Psi^{(0)}\|$ is time-independent (the time-dependent contributions to (\ref{normN}) and (\ref{normG}) cancel), and the 
ground state is stable in Friedman space-times.  
By our definition, Friedman space-times are quantum complete, albeit they are 
geodesically incomplete. 
Second, for a Schwarzschild black hole, the situation is different, because 
$g^{tt}\sqrt{|{\rm det}(g)|}W(\widehat{\varphi},\widehat{\varphi}^*) $ is 
time-dependent in this case. Hence, the ground state cannot be stable, but 
the Schwarzschild black hole can still be quantum complete (with respect to 
free fields). 

In order to show that Schwarzschild black holes are indeed quantum complete,
we transform the Riccati equation (\ref{RicattiKernel}) for the 
Fourier amplitudes $\widehat{K}$ to a homogeneous, second-order ordinary 
differential equation in normal form,
\begin{eqnarray}
	\label{normsys}
	&&\partial_\tau^{\; 2} f(k,\tau) + \omega^2(k,\tau) f(k,\tau) = 0
	\; , \\
	&& \omega^2(k,\tau) 
	\equiv
	\tfrac{r_{\rm g}^{\;2}}{16 g_{\theta\theta}(\tau)}
	\left(1 -2g_{tt}(\tau)+g_{tt}^{\;\;2}(\tau)\right) 
	\nonumber \\ 
	&& \hspace{1.6cm}- g_{tt}(\tau) M^2 \Omega^2 (k,\tau)
	\; .
\end{eqnarray}	
The Fourier amplitudes $\widehat{K}$ are related to $f$ as follows:
\begin{eqnarray}
	\widehat{K}(k,\tau)
	=
	-\tfrac{1}{2 {\rm det}(q)}\; 
	\partial_\tau \; {\rm ln}(\sigma(\tau)f^2(k,\tau))
	\; ,
\end{eqnarray}	
with $\sigma(\tau)\equiv-{\rm i}M \sqrt{|{\rm det}(g)|}/{\rm det}(q)$.

The dispersion relation for $f$ is singular at the horizon, $\tau=2$, 
and at the classical black hole singularity, $\tau=0$. For our purposes,
it suffices to expand $f$ near $\tau=0$. 
Let us first give a quick argument and justify it a posteriori.
The leading singularity in the dispersion relation around $\tau=0$
is given by $\omega_0=1/(2\tau)$, with corrections $\mathcal{O}(1/\sqrt{\tau})$.
Near $\tau=0$, the dynamics is governed by the background, i.e.~the dominant 
contribution in the dispersion relation is momentum-independent. 
In this regime, 
$f(\tau)\rightarrow C^\prime \sqrt{\tau}(C+{\rm ln}\tau)$, which translates to 
\begin{eqnarray}
	\label{domk}
	&&\operatorname{Im}\left(
	\widehat{K}(\tau)\right)\rightarrow
	\tfrac{-1}{M^3{\rm sin}(\theta)} \tfrac{1}{\tau^3|{\rm ln}\tau|}	
	\;, \nonumber \\
	&&\operatorname{Re}\left(
	\widehat{K}(\tau)\right)\rightarrow
	\left|\operatorname{Im}(C)\right| \; \tfrac{\left|\operatorname{Im}
	\left(\widehat{K}(\tau)\right)\right|}{|{\rm ln}\tau|}
	\; ,
\end{eqnarray}
near the black hole singularity. Here, $C,C^\prime\in\mathbb{C}$ are constants
of integration. Note that 
$\operatorname{Re}(\widehat{K}(\tau)) \ll \operatorname{Im}(\widehat{K})$
near the singularity. The real part is taken into account since the dominant contribution 
gives a phase factor for $\mathcal{G}^{(0)}$.
Using (\ref{domk}) in (\ref{N0}), the normalisation $N^{(0)}$ goes to zero like 
\begin{eqnarray}
	N^{(0)}(\tau) 
	\rightarrow \left|{\rm ln}\tau\right|^{-\tfrac{1}{2}v(\Sigma)\Lambda}
	\; .
\end{eqnarray}	
Of course, this evaluation 
requires a volume as well as an ultraviolet regularisation. 
We simply introduced a coordinate volume 
and an ultraviolet cut-off, $v(\Sigma)$ and $\Lambda$, respectively,
since the regularisation details have no impact on the limit
$N^{(0)}(\tau)\rightarrow 0$ as $\tau\rightarrow 0$.
For $\mathcal{G}^{(0)}$, we find  
\begin{eqnarray}
	\mathcal{G}^{(0)}(\tau)
	= 
	\exp{\left(-\tfrac{1}{2}\operatorname{Re}
	\left(\widehat{K}\right)(\tau)\int_{\Sigma_\tau}  \mbox{d}\mu(x) \; \phi^2(x)
	\right)}
\end{eqnarray}	
times an irrelevant phase factor. 

It is more rigorous to take all contributions in the dispersion relation into account 
that are singular at $\tau=0$, 
\begin{eqnarray}
	\omega_s^{\; 2}(k,\tau)
	=
	\omega_0^{\; 2} (\tau) + \left(k_\measuredangle^{\; 2}(\theta)
	+\tfrac{1}{2}\right) \omega_0(\tau)
	+\mathcal{O}(\tau^0) \; ,
\end{eqnarray}	
where $k_\measuredangle^{\; 2}\equiv (\tau r_{\rm g})^2 {\rm d}\mathfrak{s}^2(k,k)/4$.
Introducing the variable $z\equiv \sqrt{1+2k_\measuredangle^{\; 2}}\sqrt{\tau}$,
we find 
\begin{eqnarray}
	f(k,\tau) \rightarrow
	-\tfrac{z}{2}
	\left(J_0\left(z\right)
	-2{\rm i} K_0\left({\rm i}z\right)
	\right)\; ,
\end{eqnarray}	
near the black hole singularity, 
with $J_0$ denoting the Bessel function of the first kind, and 
$K_0$ denoting the modified Bessel function of the second kind.
This combination shows the same behaviour near $\tau=0$ as $f$
subject to the dispersion relation $\omega_0$. The momenta $k_\measuredangle$
in angular directions appear only in an overall factor $\ge 1$
and do not modify the dominant behaviour near $\tau=0$.

Therefore, it is safe to conclude that 
\begin{eqnarray}
	\label{result}
	\left\|\Psi^{(0)}\right\|^2(\tau)
	&\rightarrow& 
	\left|{\rm ln}(\tau)\right|^{-v(\Sigma)\Lambda}
	\left(\tau^{3/4}|{\rm ln}(\tau)|\right)^{N(\Lambda)} 
	\!\!\!\rightarrow 0\;\;\;
\end{eqnarray}	
as the black hole singularity is approached. Here, $N(\Lambda)$ denotes
the number of momentum modes with $|k|\in [0,\Lambda^{1/3}]$.
The limit (\ref{result}) is our main result. In fact, already
$\Psi^{(0)}[\phi](\tau)\rightarrow 0$ as $\tau\rightarrow 0$,
i.e.~the wave functional has vanishing support towards the singularity.

Let us stress again that we were interested in examining the quantum
completeness of Schwarzschild black holes with respect to 
free quantum fields. The answer to this question is insensitive 
to the details of volume and short-distance regularisation,
both of which are required, in principle. 

\section{Conclusion \& Discussion}
In this article we adapted the notion of quantum-mechanical completeness 
to situations where the only adequate description is in terms of a quantum 
theory of fields in generic space-times. 
We showed that according to the 
advanced consistency criterion, a Schwarzschild black hole is quantum complete
with respect to free scalar fields (in the ground state). 
Moreover, the wave functional has vanishing support towards the black hole singularity.

There are two types of non-Gaussianities that can be introduced to describe
processes associated with deviations from free fields in the ground state. 
First of all, excitations of the ground state can be considered. 
It should be clear that the term excitation is strictly appropriate for static backgrounds. 
In general, excitations will depend on eigenfunctions $\epsilon$
of $(\Delta-m^2)$ in the background geometry. 
Excited states are of the form 
$\Psi^{(n)}[\phi](\tau)=N_n[\phi,\epsilon](\tau)\Psi^{(0)}[\phi](\tau)$. So 
excitations are reflected in a (functional) renormalisation of 
$N_0(\tau)$. The difference between the ground state and the excited
states is the following: $\Psi^{(0)}[\phi](\tau)$ populates the ground state 
with field configurations that need not satisfy any on-shell criteria. 
What matters is the spatial support of the scalar fields and the  
correlation between two fields as communicated by the kernel function.  
This is why the completeness concept used here poses a rather strong 
consistency requirement on the kernel function. 
In contrast, excited states are sensitive, in addition, to the moderated 
overlap between an arbitrary field configurations and fields obeying 
on-shell conditions. Moderation indicates that the overlap is evaluated 
using $\phi(x)K(x,y,\tau)\epsilon(y)$. Intuitively, excitations show
an increasing sensitivity on the on-shell conditions. 

Secondly, interactions of the Klein-Gordon field with itself and with other fields 
can be introduced in the Hamiltonian. In this case, we choose the initial data 
such that the interactions can be treated in the usual perturbative framework. 
If the Schwarzschild black hole fails to be quantum complete with respect
to interacting fields, then the participating fields necessarily entered a strong
coupling regime, because the space-time is quantum complete with respect to free fields. 

Perhaps not surprisingly, Schwarzschild black holes are 
enjoying a clash of completeness concepts. The obvious question is how to qualify 
the importance of quantum completeness relative to classical completeness.
We think that this question is related to the measurement process. 
Let $\gamma(t)\;, t\in(0,t_0]$ be a geodesic, and 
$\{t_n\}\rightarrow 0$ denote a parameter sequence such that
$\{\gamma(t_n)\}$ does not converge.
The inextendibility of the geodesic can be observed by measuring any
classical observable 
$\mathcal{O}$ along $\gamma$: $\{\mathcal{O}(\gamma(t_n))\}\subset \mathbb{R}$ 
does not converge. Hence, geodesic incompleteness is observable, provided 
the measurement process associated with $\mathcal{O}$ is known. 
Certainly the measurement process will involve quantum theory at a more or less
obvious but essential level. 
We can ask whether the geodesic incompleteness has an impact on the quantum theory
underlying the measurement process. For instance, if black holes are quantum 
incomplete with respect to the degrees of freedom employed in the measurement device,
then $\mathcal{O}$ cannot be measured, and the geodesic incompleteness is 
not observable. If this holds for any observable, then the geodesic incompleteness 
is unobservable in principle. This may sound impractical as a criterion. 
Measurement processes, however, rely on a few principles and are realised 
via universal principles such as minimal coupling. 
This makes it relatively easy to pass from unobservable to observable in principle. 

We found that Schwarzschild black holes are quantum complete, and, moreover,
the ground state does not support field configurations near the singularity.
The logical conflict with the measurement process as described above has a well-known 
resolution: Near the black hole singularity, observables necessarily are part and parcel
of the quantum theory.
So consistency of the quantum theory is not only essential for the measurement device,
but already for the very construction of observables. 

In our opinion, and in conclusion, the concept of quantum completeness as suggested 
in this work has physical relevance, and presents a physical characterisation
of space-time singularities and their impact.


\begin{acknowledgments}
It is a great pleasure to thank Cesar Gomez, 
Andre Franca, Sophia M\"uller, Florian Niedermann, 
Tehseen Rug and Robert Schneider for delightful discussions, and 
Kristina Giesel and Thomas Thiemann for their thoughts on the topic 
during a cold night. 	
We appreciate financial support of our work
by the DFG cluster of excellence 'Origin and Structure of the Universe'
and by TRR 33 'The Dark Universe'. 

\end{acknowledgments}





\bibliography{clvsqmcompl}

\end{document}